\documentclass[aps,superscriptaddress, twocolumn,a4paper]{revtex4}

\usepackage{hyperref}
\usepackage[sort&compress]{natbib}
\usepackage{latexsym}
\usepackage{epsfig}
\usepackage[english]{babel}
\usepackage{graphicx}
\usepackage[T1]{fontenc}
\usepackage[utf8]{inputenc}

\usepackage{amsmath,amsfonts,amssymb}

\newcommand \be {\begin{equation}}
\newcommand \bea {\begin{eqnarray}}
\newcommand \ee {\end{equation}}
\newcommand \eea {\end{eqnarray}}

\newcommand \bed {\begin{displaymath}}
\newcommand \eed {\end{displaymath}}

\newcommand{\bit}{\begin{itemize}}
\newcommand{\eit}{\end{itemize}}

\newcommand{\bgar}{\begin{eqnarray}}
\newcommand{\enar}{\end{eqnarray}}

\begin{document}

\title{Containment effort reduction and regrowth patterns of the Covid-19 spreading}

\author{D. Lanteri}
\affiliation{INFN, Sezione di Catania, I-95123 Catania, Italy}
\affiliation{Dipartimento di Fisica e Astronomia, Universit\`a di Catania, Italy}

\author{D.~Carco'}
\affiliation{Istituto Oncologico del Mediterraneo, Viagrande, Italy}

\author{P.~Castorina}\email[]{Corresponding author}
\affiliation{INFN, Sezione di Catania, I-95123 Catania, Italy}
\affiliation{Faculty of Mathematics and Physics, Charles University, V Hole\v{s}ovi\v{c}k\'ach 2, 18000 Prague 8, Czech Republic} 

\author{M.Ceccarelli}
\affiliation{U.O.C. Malattie Infettive, P.O. Garibaldi, Catania, Italy}

\author{B.Cacopardo}
\affiliation{U.O.C. Malattie Infettive, P.O. Garibaldi, Catania, Italy}
\affiliation{Dipartimento di Medicina clinica e sperimentale, Universit\`a di Catania, Italy}


\date{\today}

\begin{abstract}
\noindent In all Countries the political decisions aim to achieve an almost stable configuration with a small number of new infected individuals per day due to Covid-19.  When such a condition is reached, the containment effort is usually reduced in favor of a gradual reopening of the social life and of the various economical sectors. However, in this new phase, the infection spread restarts and a quantitative analysis of the regrowth is very useful.  We discuss a macroscopic approach which, on the basis  of the collected data in the first lockdown, after few days from the beginning of the new phase, outlines different scenarios of the Covid-19 diffusion for longer time.  The purpose of this paper is a demonstration-of-concept: one takes simple growth models, considers the available data   and shows how the future trend of the spread can be obtained. The method applies a time dependent carrying capacity, analogously to many macroscopic growth laws in biology, economics and population dynamics. The illustrative cases of Singapore, France, Spain and Italy are analyzed. 
\vskip 10pt
{\bf keywords:} Covid-19 spreading, mathematical models, macroscopic growth laws, carrying capacity 

\end{abstract}
\maketitle

\section{Introduction}

The pandemic spreading of the  Coronavirus infection  2019 (COVID-19)~\cite{oms,who,hopkins} is forcing billion of people to live in isolation. The related economical degrowth is producing dramatic conditions for workers, trade and industry.

In all Countries the political decisions aim to reduce the spreading  and to achieve an almost stable configuration of coexistence with the disease, where a small number of new infected individuals per day is sustainable. In this new  condition, the containment effort is usually reduced in favor of a gradual reopening of the social life and of the various economical sectors: the so called phase 2 (Ph2).  

In the Ph2 the spread usually restarts and  the evaluation of the regrowth of the infection diffusion is a complex problem: microscopic models require a coupled dynamics of the stakeholders, implying a strong model dependence and  a large number of free parameters~\cite{napoco1,napoco2,napoco3,napoco4,pluc, epid1}.
For example, the asymptomatic population has been estimated about $\le 50$ \cite{imp}, $\simeq 10$ \cite{istat}, $\simeq 3 -4 $ \cite{lancet, noib} times the symptomatic one and
the simulation in the Italian report on the effects of the reopening on the National Health System is based on a stochastic epidemic model including the age dependence, the demographic structure, the heterogeneity of social contacts in different meeting places (home, school, work, public transportation, cultural activity, shop, bank, post office) and  many work sub-sectors (public health, manufacturing, building, trade,~...)~\cite{gov}. 

On the other hand, complementary approaches, which outline the Covid-19 evolution in Ph2 in a model independent way, on the basis of macroscopic growth laws (with few parameters)  are a useful tool for monitoring the regrowth of the spreading by collecting data after few days from the end of the lockdown or, in general, of the restarting phase. 

In this paper we propose a method, based on macroscopic variables~\cite{noi1,noi2,noia,noib} and with no explicit reference to  the underlying dynamics, which analyzes the quantitative consequences of the impairment of the constraints.

The starting point is the observation that the Covid-19 spreading, after an initial exponential increase and a subsequent small slowdown (which follows the Gompertz law (GL)~\cite{gompertz} or other non linear trends),  reaches a saturation, stable phase, described by the GL or by a logistic equation (LL)~\cite{logistic}, after which the Ph2 starts. 

The GL, initially applied to human mortality tables (i.e. aging), also describes tumor growth, kinetics of enzymatic reactions, oxygenation of hemoglobin, intensity of photosynthesis as a function of CO2 concentration, drug dose-response curve, dynamics of growth, (e.g., bacteria, normal
eukaryotic organisms). The LL~\cite{logistic} has been used in population dynamics, in economics, in material science and in many other sectors.

The previous macroscopic growth laws, GL and LL, depend on two parameters, related to the initial exponential trend and to the maximum number of infected individuals, $N_\infty$, called carrying capacity.

It is well known that the carrying capacity changes according to some ``external'' conditions in many biological, economical and social systems~\cite{review2}.
In tumor growth it is related to a multi-stage evolution~\cite{wehldon}. In population dynamics, new technologies affect how resources are consumed, and since the carrying capacity depends on the availability of that resource, its value changes~\cite{pop}. 

Therefore a simple method of monitoring the Ph2 is to understand how the carrying capacity (CC) increases due to the reduction of the social isolation and to the restarting of the economical activities. As discussed, this modification is difficult to predict, but different scenarios of regrowth (i.e. with different time dependence of CC in the Ph2, for example) are analyzed in the next sections.

 By monitoring the  initial data in the new phase one outlines the behavior of the spreading for longer time to evaluate the possible effects of new mobility constraints or new total lockdown.
If the infection regrows exponentially, the (re)lockdown and/or other containment efforts have to be decided as soon as possible. On the other hand, a small change of the specific rate in the Ph2, parameterized by a slight modification of the CC, should require less urgent political choices.

\section{\label{sec:1} Theory and calculations}

\subsection{Macroscopic growth law with time dependent carrying capacity}

The macroscopic growth laws for a population $N(t)$ are solutions of a general differential equation that can be written as
\be
\frac{1}{N(t)} \frac{dN(t)}{dt} = f[N(t)]
\ee
where $f(N)$ is the specific growth rate and its $N$ dependence describes the feedback effects during the time evolution. If $f(N)=$~constant, the growth follows an exponential pattern.

In particular, the Gompertz  and the logistic equations are
\be\label{eq:G}
\frac{1}{N(t)}\frac{dN(t)}{dt} = - k_g\;\ln \frac{N(t)}{N_\infty^g}  \qquad \text{Gompertz}\;,
\ee
\be\label{eq:L}
\frac{1}{N(t)}\frac{dN(t)}{dt} =  k_l \left(1- \frac{N(t)}{N_{\infty}^l}\right)
\qquad 
\text{logistic},
\ee
where  $k_g \ln(N_\infty^g)$ and $k_l$ are respectively the initial exponential rates and the other terms determine their slowdown. In both cases the steady state condition, $dN/dt =0$ is reached when $N$ is equal to the carrying capacity $N_\infty$.

As shown in refs. \cite{noib,noia}  the coronavirus spreading has, in general, three phases: an initial exponential behavior, followed by a Gompertz one and a final logistic phase, due to lockdown. 

In many dynamical systems the previous, simple, GL or LL  solutions give a good quantitative understanding of the growth. However the CC can be modified by effects not included in eqs.~(\ref{eq:G},~\ref{eq:L}). For example, the invention and diffusion of technologies  lift the growth limit.

For Covid-19 infection, in the Ph2 phase there is a fast increase of the human mobility  and aggregation  which, considering the large number of asymptomatic individuals,  modifies the CC.
Moreover, one has to take into account that other pathological features of the virus could be different in the new phase (viral load, external temperature, ...) also.

Therefore one introduces an extension to the widely-used macroscopic model to allow for a time dependent carrying capacity and a change in the parameter which characterize the exponential initial phase, due to
the different infectious features of the Covid-19. In other terms, eqs.~(\ref{eq:G},~\ref{eq:L}) are now 
respectively coupled with different values of the constant $k_g$ or $k_l$ and  a differential equation for the evolution of the CC, i.e. (g=Gompertz, l=logistic)
\be\label{eq:CC}
\frac{d N_\infty^{g,l}}{dt} = \beta^{g,l}(t) 
\ee
where $\beta^{g,l}(t)$ are the rates: $\beta=0$, $\beta=$constant, $\beta \simeq t^n$, $\beta \simeq c\;\exp(b\;t)$ give respectively constant, linear, power law and exponential time dependence of the CC.

\subsection{Covid-19 spreading in phase 2 - formulation}

The application of the previous differential equations to the spreading of Covid-19 in the Ph2 in different Countries requires: a) the time, $t^\star$, of the beginning of the change  of the isolation conditions and/or of the restarting phase;
b) a stable phase of the infection diffusion for $t < t^\star$: the effects of the political decision of reducing the constraints start (or should start) when the disease shows a clear slowdown (see below). 

Therefore for $t \le t^\star$ the total number of infected individuals is described by eqs.~(\ref{eq:G},\ref{eq:L}) with constant $N_\infty^{g,l}$ fitted by the available data, and for $t \ge t^\star$ one has to solve the system of coupled differential equations~(\ref{eq:G}-\ref{eq:CC})   where the CC is a function of time, with the initial condition that $N_\infty^{g,l}(t^\star)= N_\infty^{g,l}$.

A simple example is useful to outline the strategy. If a time $t^*$ the spread is stable, then $N(t^*) \simeq N_{\infty}^{(g,l)}$ and the specific rate is very small. Let us assume tha for $t>t^*$ there is a fast
rate of the spreading which follows the GL in the new phase with a new CC, i.e.
\be
\frac{1}{N(t)}\frac{dN(t)}{dt} = - k_g\;\ln \frac{N(t)}{N_\infty^{(g2)}} \phantom{....} for \phantom{....} t > t^*\;,
\ee
where $N_\infty^{(g2)} > N_\infty^g$ is the carrying capacity in the new phase and $N(t^*)$ is the initial value of the regrowth. If $N_\infty^{(g2)}= \gamma  N_\infty^g$, with constant $\gamma$, the Gompertz equation for $t>t^*$ is given by
\be
\frac{1}{N(t)}\frac{dN(t)}{dt} = - k_g\;\ln \frac{N(t)}{N_\infty^{(g2)}}= - k_g\;\ln \frac{N(t)}{\gamma N_\infty^g}
\ee
that is
\be
\frac{1}{N(t)}\frac{dN(t)}{dt} = + k_g\;\ln \gamma - k_g\;\ln \frac{N(t)}{\gamma N_\infty^g}
\ee
and if $\ln \gamma >> \ln [N(t^*)/N_\infty^g]$ a new exponential phase of the spreading starts for $t > t^*$.

\begin{figure}
	\includegraphics[width=\columnwidth]{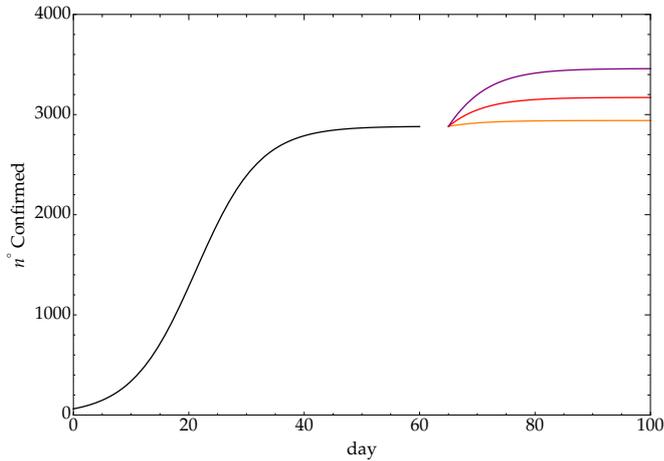}
	\caption{Variation of a logistic growth due to a sudden change in the CC: $N_\infty^{l\,\star} = k\;N_\infty^l$, with $k=1.02$ (orange), $k=1.1$ (red) and $k=1.2$ (purple).}
	\label{fig:4}
\end{figure}

The condition $t>t^\star$ has to be better clarified. The instantaneous change of the CC is unphysical since there is a time interval to observe a possible increase of the spreading due to the Covid-19  incubation time, $\Delta$. Therefore in the time interval $t^\star < t < t^\star + \Delta $ the growth behavior still follows the initial phase, with a fixed CC.  
The study of the incubation time is crucial to define the delay  (after $t^\star$) of  a possible regrowth. This aspect is discussed in the next section
and to clarify the proposed method let us assume that a logistic trend up to $t^\star=60$ days, with a CC, $N_\infty^l= 2883$, is modified at the day $t^\star=60 + \Delta$, with $\Delta=5$,  by an increase of the CC by a constant factor ($1.02$, $1.1$, $1.20$). Fig.~\ref{fig:4} shows the cumulative number of detected infected individuals. 

The previous examples are for illustrative purposes and  in the next sections we apply the proposed approach to Singapore, France, Spain and Italy, including the effects of the delay $\Delta$.

\subsection{Covid-19 incubation time}

The definition of the incubation time, or the time from infection to illness onset, is necessary to inform choices of quarantine periods, active monitoring, surveillance, control and modeling. COVID-19 emerged just recently, and the presence of a high rate of asymptomatic individuals, does not currently allow a precise estimation of incubation time.
Different studies, especially at the beginning of the pandemic, tried to define the incubation period, obtaining a mean time varying between 4.0 and 6.4 days~\cite{baker,guan,li}.

This value of incubation time is similar to other Coronaviruses, such as MERS-CoV and SARS-CoV, and generally accepted as a reliable estimate.
However, $95\%$ confidence intervals are large, varying from 2.4 days to 15.5 days~\cite{baker}. This strong variability is related to an uncertainty of the most probable date of exposure and onset of symptoms and this is the main reason why the  WHO recommended an isolation time of 14 days after exposure to avoid more spreading of the infection~\cite{lei}.

In our study, knowledge of the incubation time is necessary to model possible consequence of a re-opening. As a matter of fact, reduction of social isolation will increase the CC, and our attention should still be at its highest levels for at least two entire incubation periods, to promptly recognize any warning signal and apply the right control measures.

\begin{figure}
	\includegraphics[width=\columnwidth]{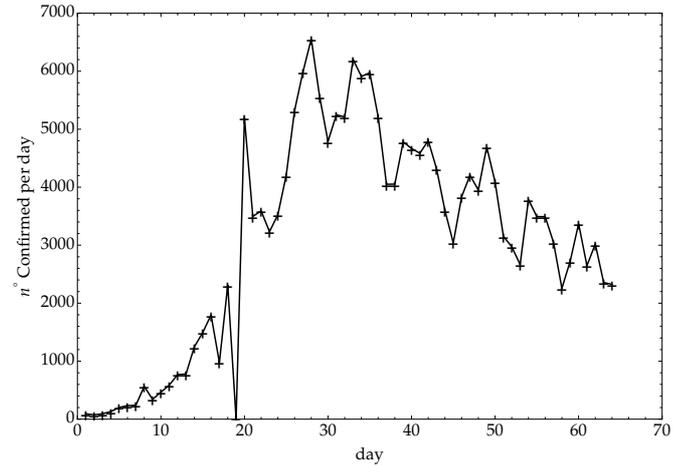}
	\caption{Italy - Confirmed daily rate.}
	\label{fig:ConfDay}
\end{figure}

\begin{figure}
	\includegraphics[width=\columnwidth]{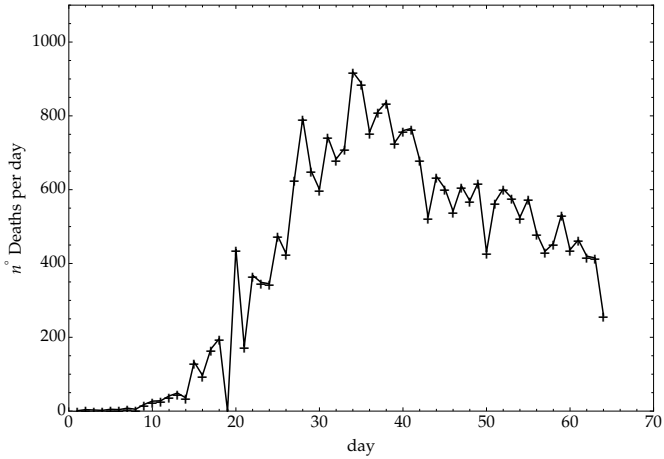}
	\caption{Italy - Mortality daily rate.}
	\label{fig:DDay}
\end{figure}

Therefore the incubation time  $\Delta \simeq 8 \pm 6 $ days can be considered  and $\Delta=6$ will be used in the next sections. Let us recall that an increase of Covid-19 mortality in Ph2 should be observed after a longer time interval. In Italy, for example, the correlation between the rate of infected people per day and the corresponding mortality rate shows a delay of about 8 days (see figs.~\ref{fig:ConfDay} and \ref{fig:DDay}). Therefore an increase of mortality could be expected after 14-22 days from $t^\star$. 

\section{\label{sec:4} Results and Discussion}

The analysis of the regrowth phase has been done with three possible trends:
the new phases is described by a LL/GL with carrying capacity 
\begin{itemize}
	\item[a)]  
	$N_\infty^{(2)} = \gamma\; N_\infty^{(1)}$,	
	
	\item[b)] $N_\infty^{(2)} =  N_\infty^{(1)} + \gamma\;\left(t-t_0\right)$,
	
	\item[c)] $N_\infty^{(2)} =  N_\infty^{(1)} \; e^{\gamma\;\left(t-t_0\right)}$.
\end{itemize}
In the next sections, $G^{a,b,c}$ and $L^{a,b,c}$ indicate the fits and the time evolution with the GL and LL, respectively, in the corresponding case $a,b,c$.

\subsection{Singapore: an early case of regrowth}

\begin{figure}
	\includegraphics[width=\columnwidth]{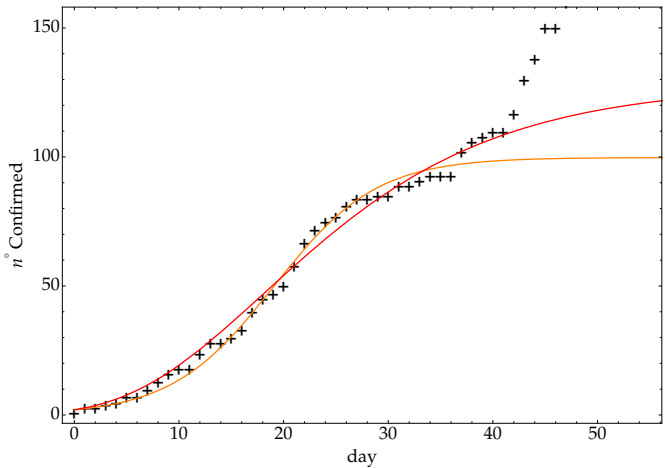}
	\caption{Singapore before the restarting of the infection. GL (red) and Logistic (orange) fit are plotted. Time zero corresponds to the initial day - 23/01.}
	\label{fig:Singapore}
\end{figure}

In Singapore, after reaching a stable phase, a new strong growth of the infection spreading has been observed, due to the immigration of workers from neighboring Countries.
This effect can be described in terms of a modified CC with respect to the saturation phase. The data of $N(t)$ before the restart of the infection can be fitted either by Gompert (red line) or by logistic (orange line), as shown in fig.~\ref{fig:Singapore}. The initial day of the PH2, $t^\star$, corresponds to about $t=32$ and $t^*+\Delta =38$ (March 2). 

\begin{figure}
	\includegraphics[width=\columnwidth]{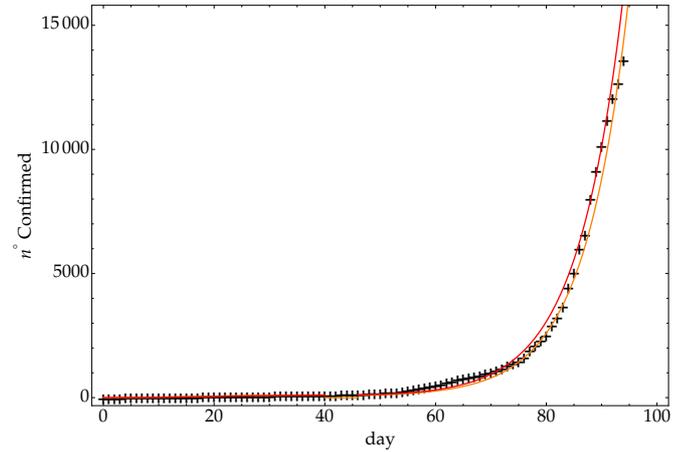}
	\caption{Singapore: GL (red) and Logistic (orange) fit with an exponential grow for $N_\infty^{g,l}(t)$ are plotted. Time zero corresponds to the initial day - 23/01.}
	\label{fig:Singapore2}
\end{figure} 

By applying the method discussed in the previous section, the entire data sample can be fitted by assuming that the CC in the new phase, $N_\infty^{g,l}(t)$,
has an exponential growth, as shown in fig.~\ref{fig:Singapore2}.

The indication coming from the previous analysis is that the increase of the spreading rate, observed immediately after the starting of the new phase, 
is so strong to require a sudden (re)lockdown of the Country.

\section{ Phase 2 in Italy: possible scenarios}

In the first phase, the Italian data  followed a GL. 
Recently, the Ph2 phase started in early September, and
different regrowth scenarios will be outlined by assuming
an increase of the CC.

\begin{figure}
	\includegraphics[width=\columnwidth]{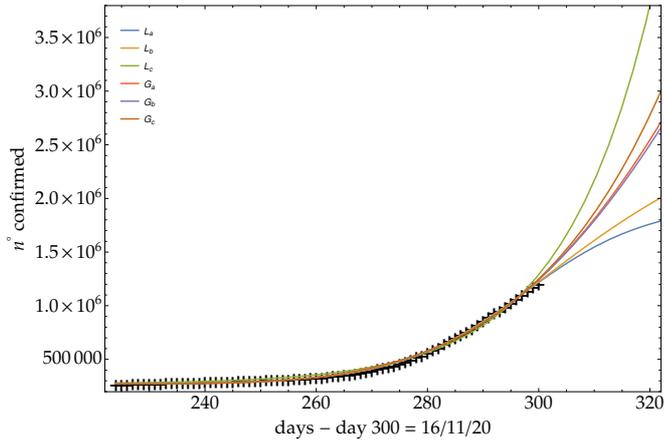}
	\caption{Italy: comparison of the growth laws with the data of the cumulative number of confirmed infected individuals from September the 1st to the final day in figure.}
	\label{fig:ItalyC}
\end{figure}

\begin{figure}
	\includegraphics[width=\columnwidth]{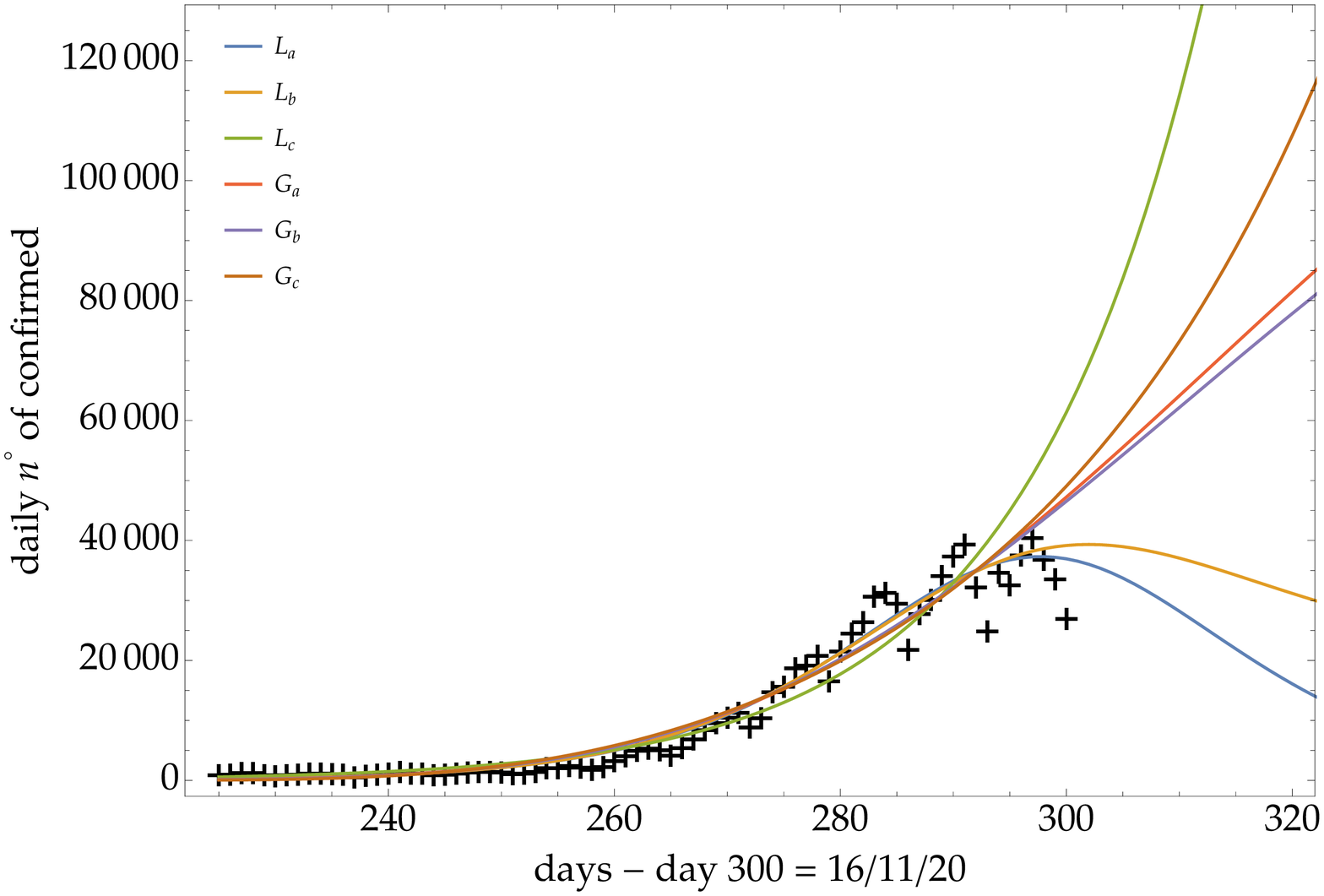}
	\caption{Italy: comparison of the growth laws with the data of the daily number of confirmed infected individuals from September the 1st to the final day in figure.}
	\label{fig:ItalyCday}
\end{figure}

\begin{figure}
	\includegraphics[width=\columnwidth]{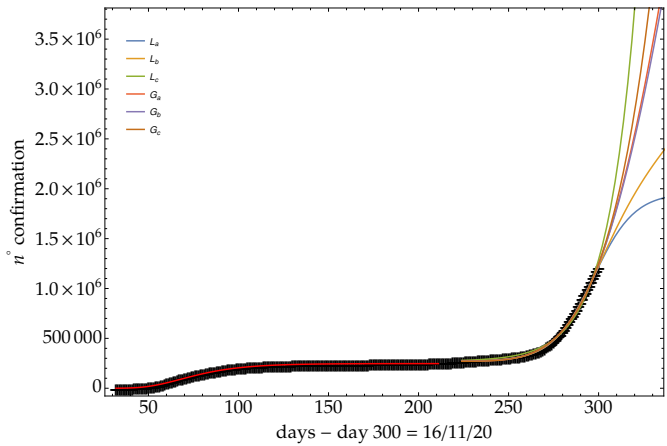}
	\caption{Italy: comparison of the growth laws with the data of the cumulative number of confirmed infected individuals from the initial day 22/Feb to the final day in figure.}
	\label{fig:ItalyCtot}
\end{figure}

For the previous trends (a,~b,~c), figures~\ref{fig:ItalyC}, \ref{fig:ItalyCday} and~\ref{fig:ItalyCtot}  show respectively the comparison of the growth laws   with the data of the cumulative number of confirmed infected individuals from $t^* + \Delta=$ September the 1st to the final day, of the daily number of confirmed infected individuals in the same period, and  of the daily number of confirmed infected individuals from the initial day 22/Feb to the final day, respectively.

\begin{figure}
	\includegraphics[width=\columnwidth]{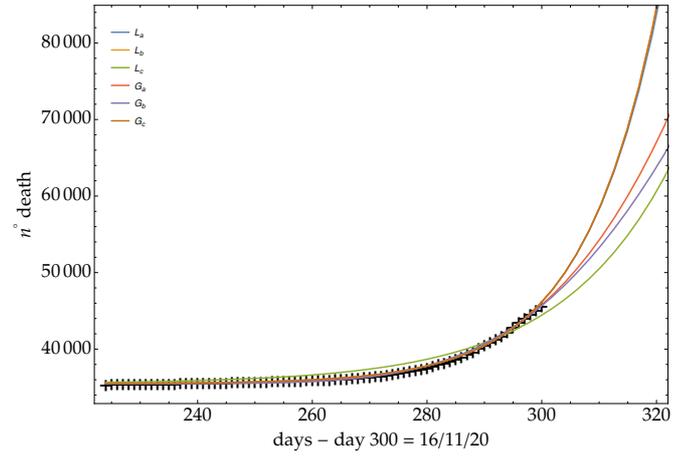}
	\caption{Italy: comparison of the growth laws with the data of the cumulative number of deaths from September the 1st to the final day in figure.}
	\label{fig:ItalyD}
\end{figure}

\begin{figure}
	\includegraphics[width=\columnwidth]{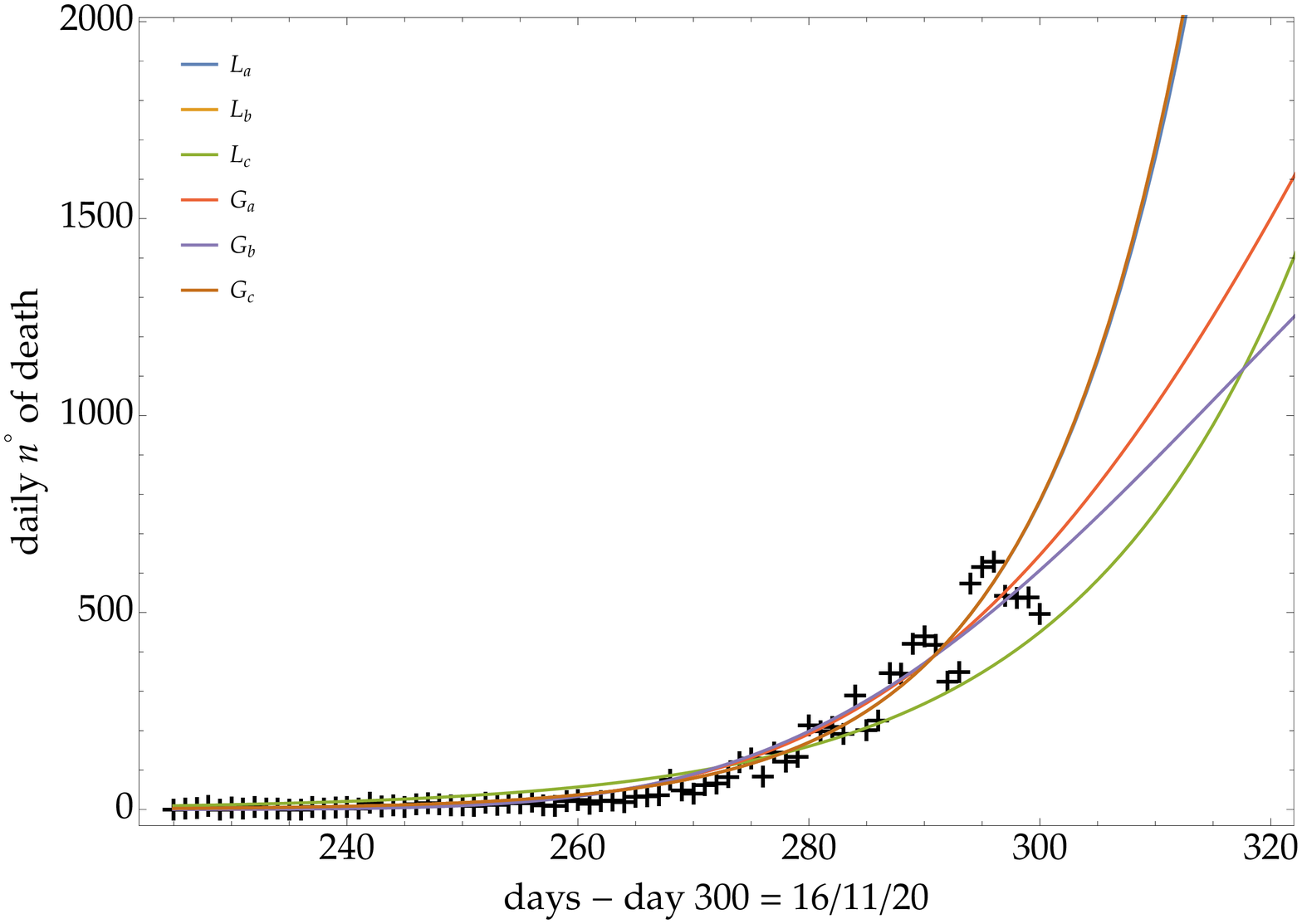}
	\caption{Italy: comparison of the growth laws with the data of the daily number of deaths from September the 1st to the final day in figure.}
	\label{fig:ItalyDday}
\end{figure}

\begin{figure}
	\includegraphics[width=\columnwidth]{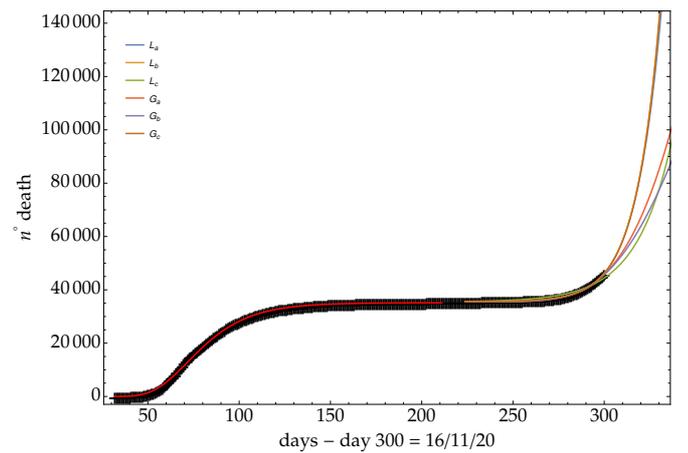}
	\caption{Italy: comparison of the growth laws with the data of the cumulative number of deaths from the initial day 22/Feb to the final day in figure.}
	\label{fig:ItalyDtot}
\end{figure}

Figures~\ref{fig:ItalyD}, \ref{fig:ItalyDday} and~\ref{fig:ItalyDtot} depict the analogous comparisons for the cumulative number of deaths, for the daily number of deaths and for  the daily number of deaths from the initial day 22/Feb to the final day, respectively.

The effects of the mobility constraints decided by the Italian government can be monitored by looking at the different predicted trends.

\section{\label{sec:FRANCE} France}

France is in a strong spread of the virus, started in August 2020, with an almost total lockdown.

\begin{figure}
	\includegraphics[width=\columnwidth]{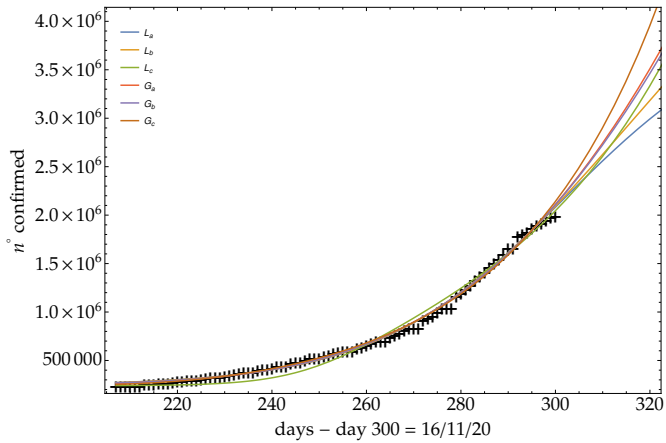}
	\caption{France: comparison of the growth laws with the data of the cumulative number of confirmed infected individuals from the 15th of August  to the final day in figure.}
	\label{fig:FranceC}
\end{figure}

\begin{figure}
	\includegraphics[width=\columnwidth]{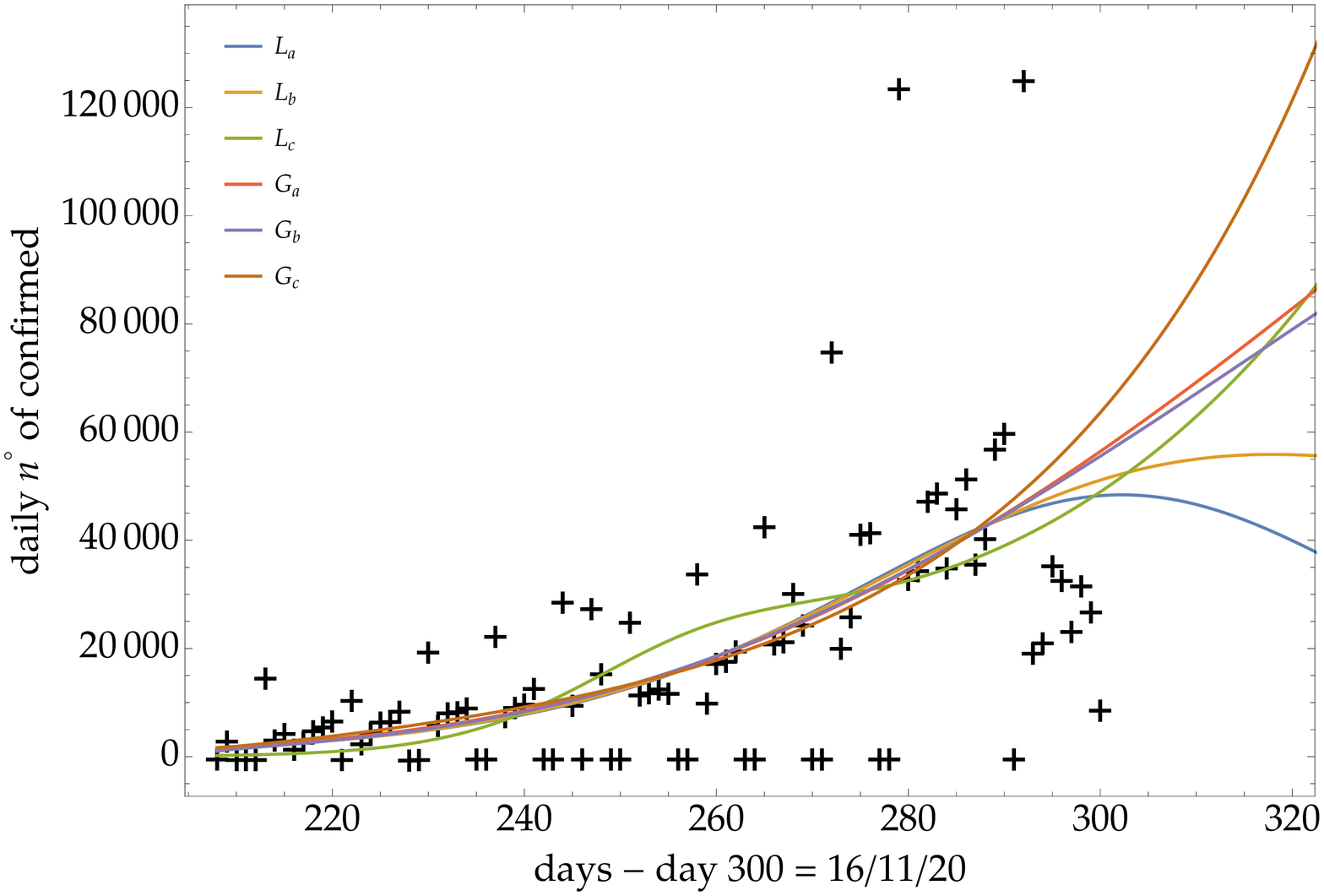}
	\caption{France: comparison of the growth laws with the data of the daily number of confirmed infected individuals from the 15th of August to the final day in figure.}
	\label{fig:FranceCday}
\end{figure}

\begin{figure}
	\includegraphics[width=\columnwidth]{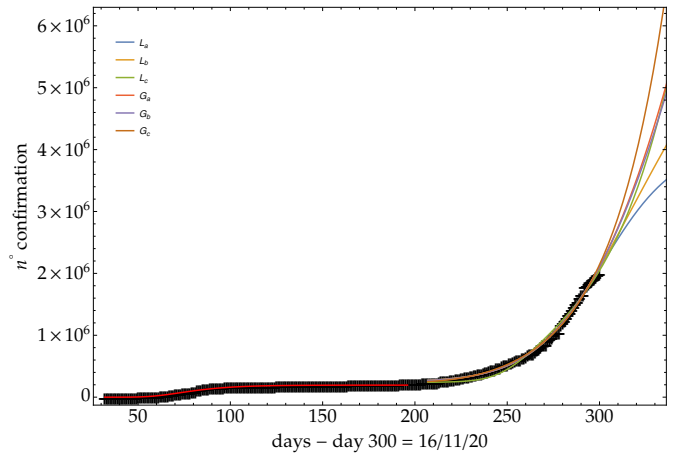}
	\caption{France: comparison of the growth laws with the data of the cumulative number of confirmed infected individuals from the initial day 22/Feb to the final day in figure.}
	\label{fig:FranceCtot}
\end{figure}

As in the previous case, figures~\ref{fig:FranceC}, \ref{fig:FranceCday} and~\ref{fig:FranceCtot}  are devoted to  the comparison of the growth laws with the data of the cumulative number of confirmed infected individuals from the 15th of August to the final day, of the daily number of confirmed infected individuals in the same period and  of the daily number of confirmed infected individuals from the initial day 22/Feb to the final day, respectively.

\begin{figure}
	\includegraphics[width=\columnwidth]{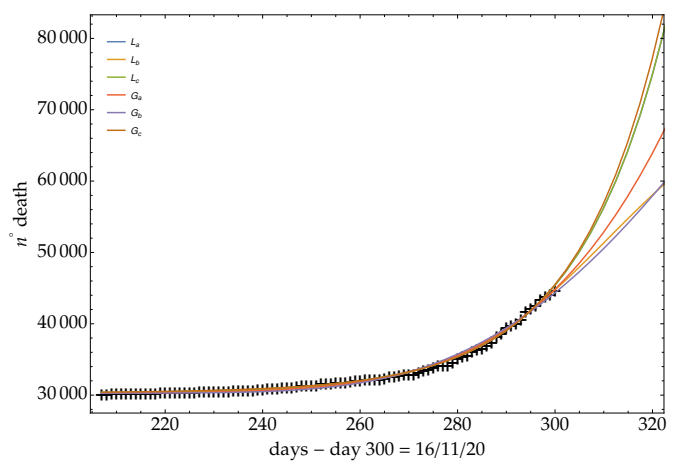}
	\caption{France: comparison of the growth laws with the data of the cumulative number of deaths from the 15th of August  to the final day in figure.}
	\label{fig:FranceD}
\end{figure}

\begin{figure}
	\includegraphics[width=\columnwidth]{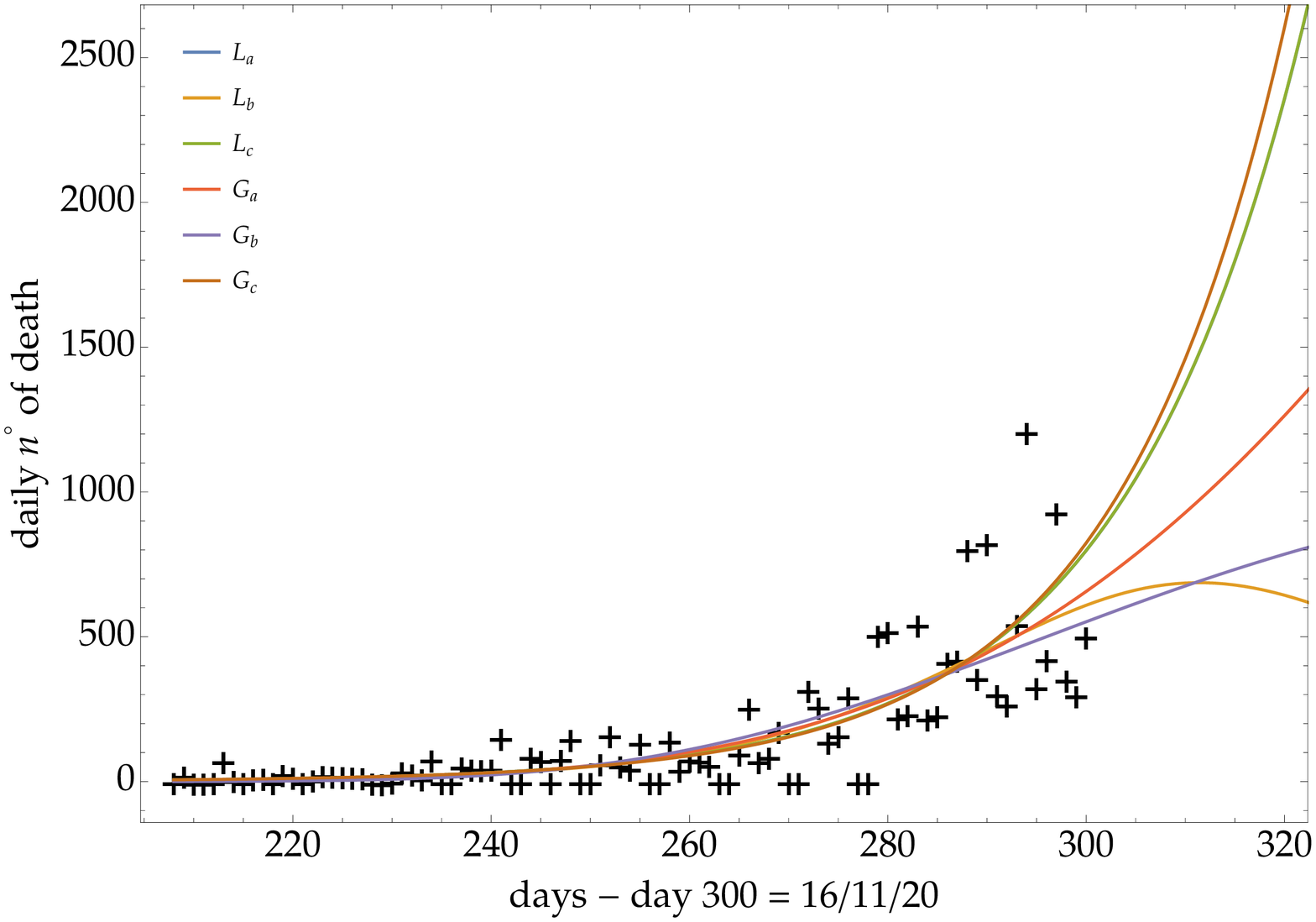}
	\caption{France: comparison of the growth laws with the data of the daily number of deaths from the 15th of August to the final day in figure.}
	\label{fig:FranceDday}
\end{figure}

\begin{figure}
	\includegraphics[width=\columnwidth]{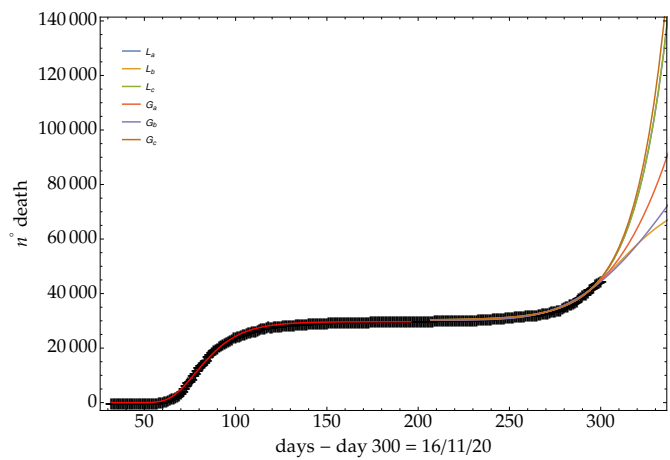}
	\caption{France: comparison of the growth laws with the data of the cumulative number of deaths from the initial day 22/Feb to the final day in figure.}
	\label{fig:FranceDtot}
\end{figure}

In figures~\ref{fig:FranceD}, \ref{fig:FranceDday} and~\ref{fig:FranceDtot}  the number of deaths for the same time periods  are reported.

\section{\label{sec:SPAIN} Spain}

\begin{figure}
	\includegraphics[width=\columnwidth]{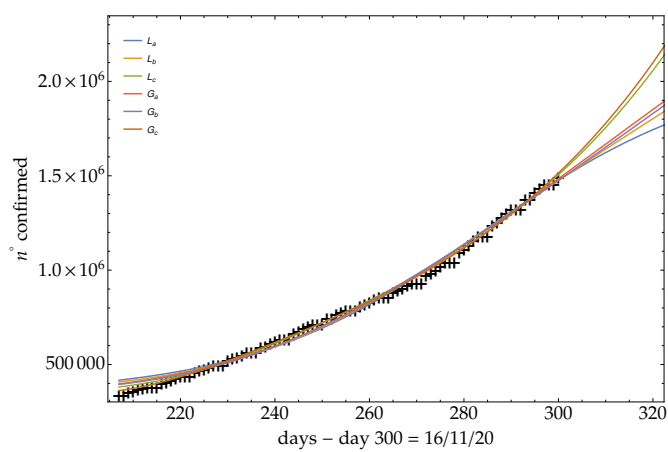}
	\caption{Spain: comparison of the growth laws with the data of the cumulative number of confirmed infected individuals from the 15th of August  to the final day in figure.}
	\label{fig:SpainC}
\end{figure}

\begin{figure}
	\includegraphics[width=\columnwidth]{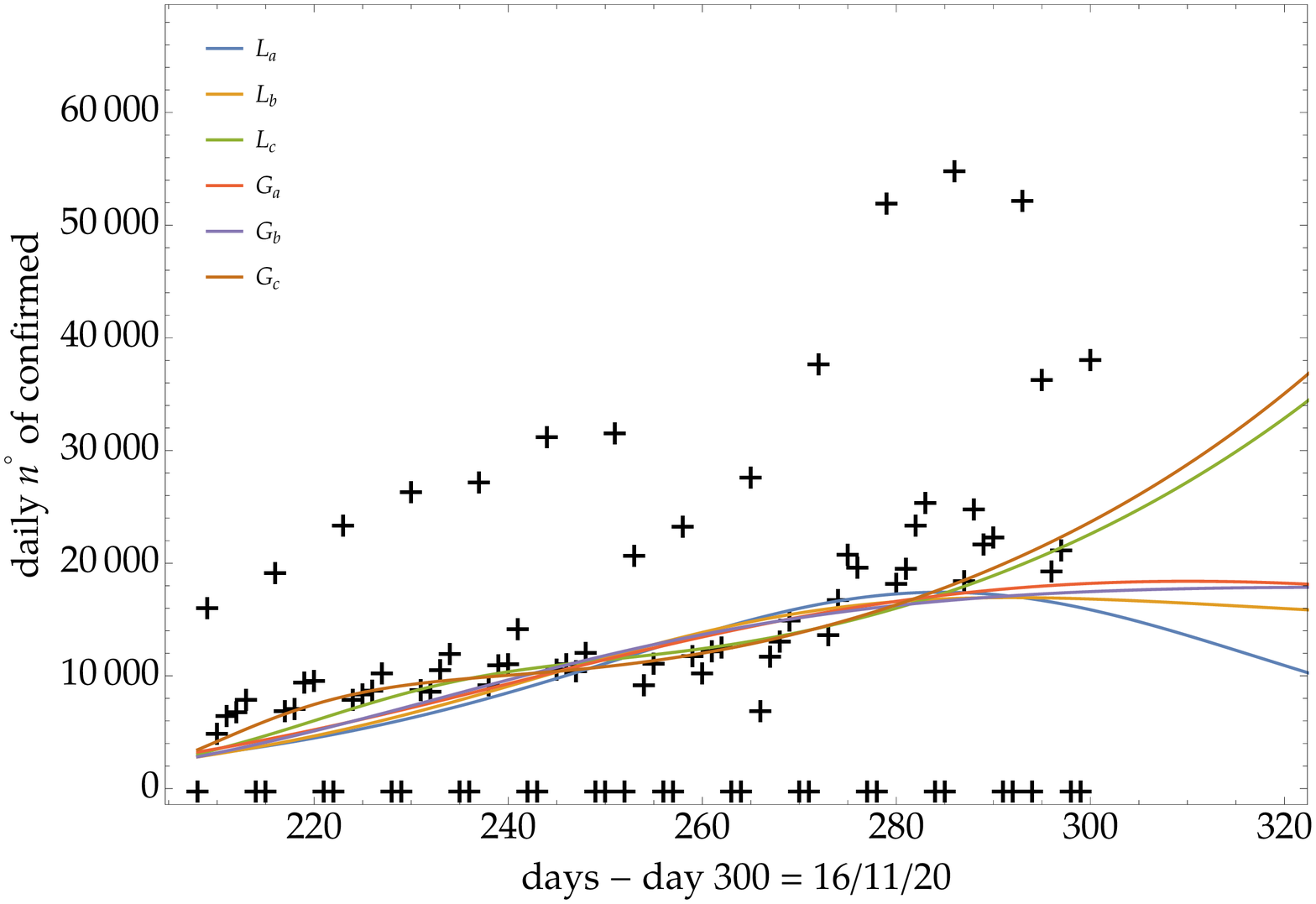}
	\caption{Spain: comparison of the growth laws with the data of the daily number of confirmed infected individuals from the 15th of August to the final day in figure.}
	\label{fig:SpainCday}
\end{figure}

\begin{figure}
	\includegraphics[width=\columnwidth]{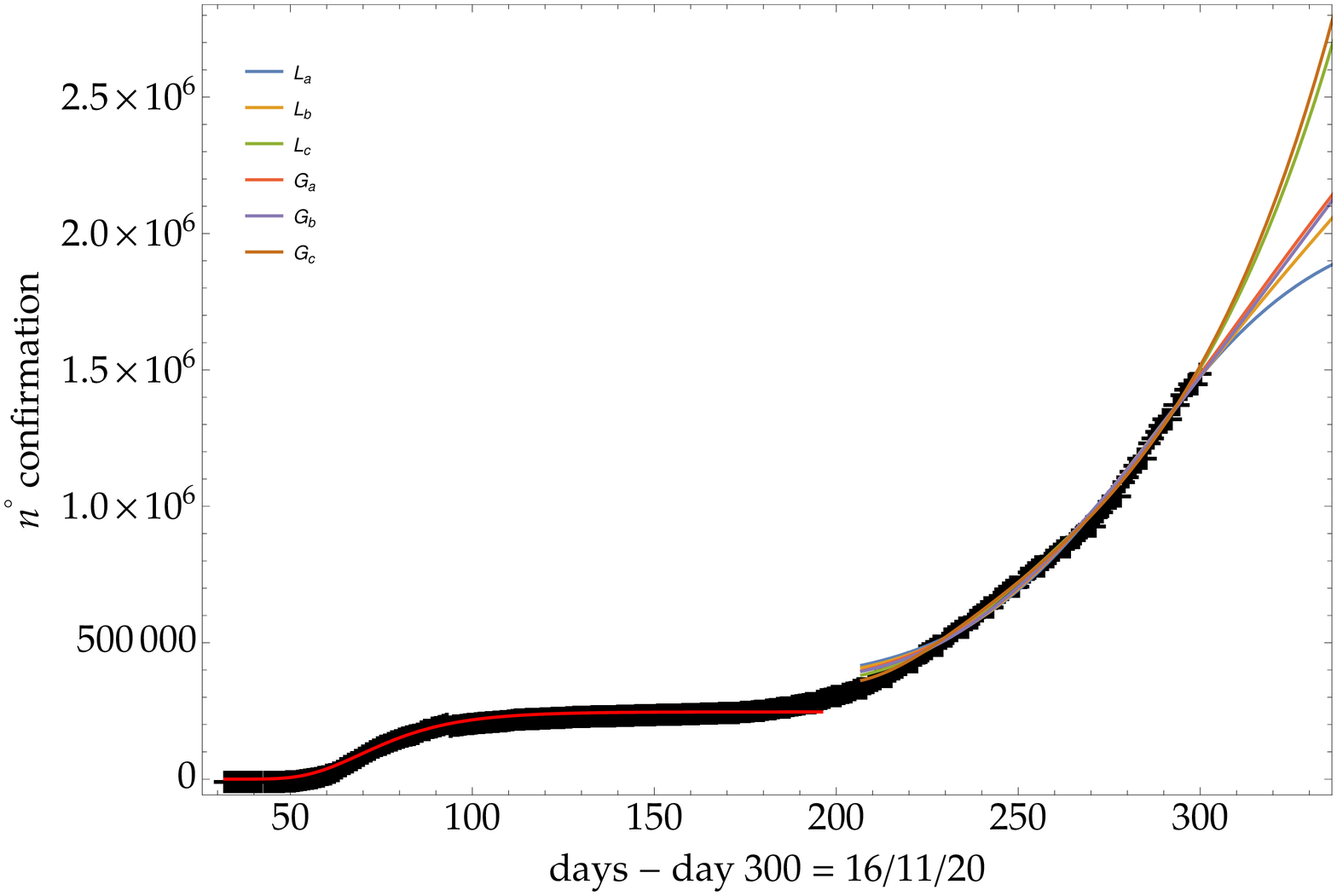}
	\caption{Spain: comparison of the growth laws with the data of the cumulative number of confirmed infected individuals from the initial day 22/Feb to the final day in figure.}
	\label{fig:SpainCtot}
\end{figure}

Figures~\ref{fig:SpainC}, \ref{fig:SpainCday} and~\ref{fig:SpainCtot} compare the growth laws with the data of the cumulative number of confirmed infected individuals from the 15th of August to the final day, of the daily number of confirmed infected individuals in the same period, and  of the daily number of confirmed infected individuals from the initial day 22/Feb to the final day, respectively.

\begin{figure}
	\includegraphics[width=\columnwidth]{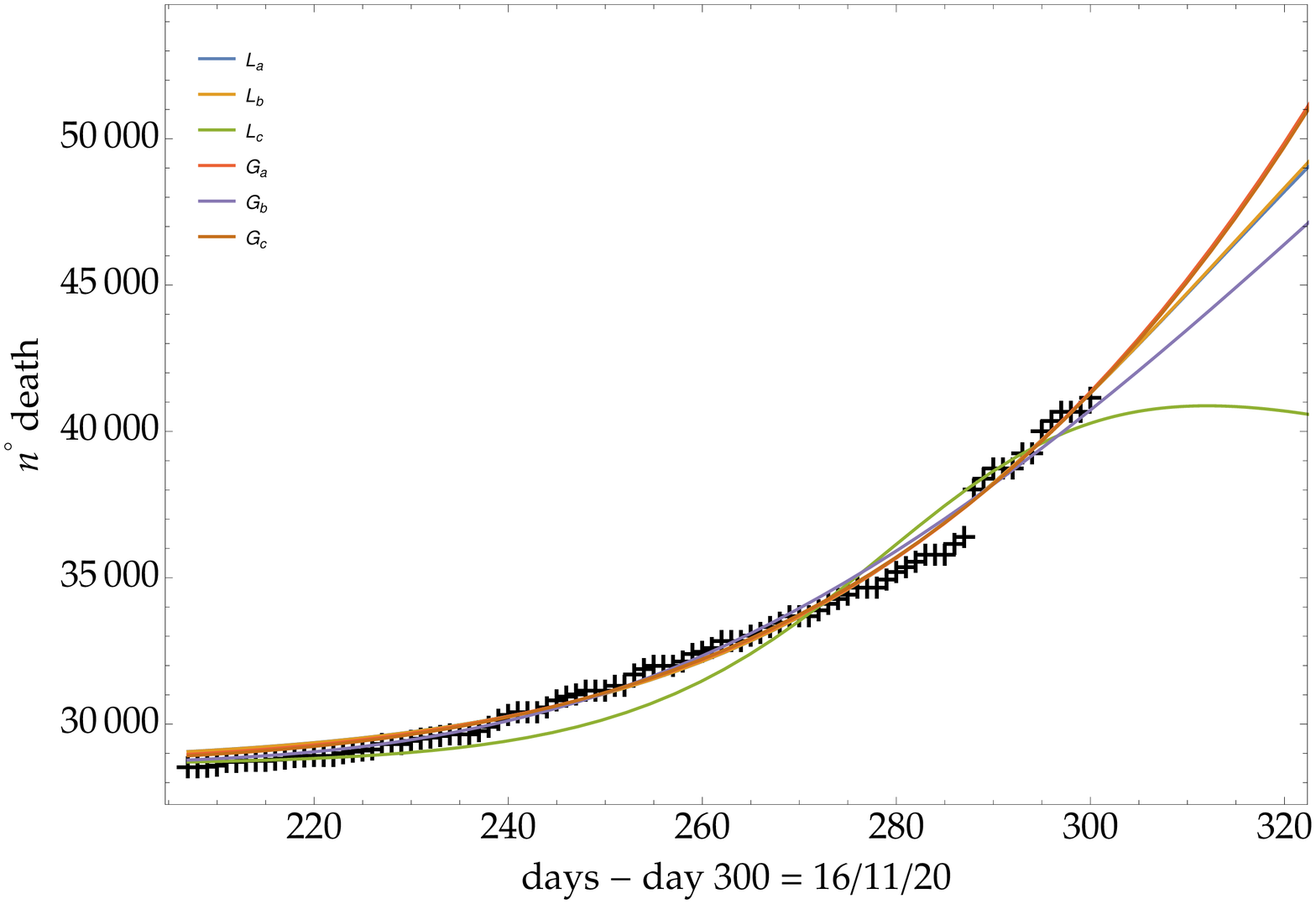}
	\caption{Spain: comparison of the growth laws with the data of the cumulative number of deaths from the 15th of August  to the final day in figure.}
	\label{fig:SpainD}
\end{figure}

\begin{figure}
	\includegraphics[width=\columnwidth]{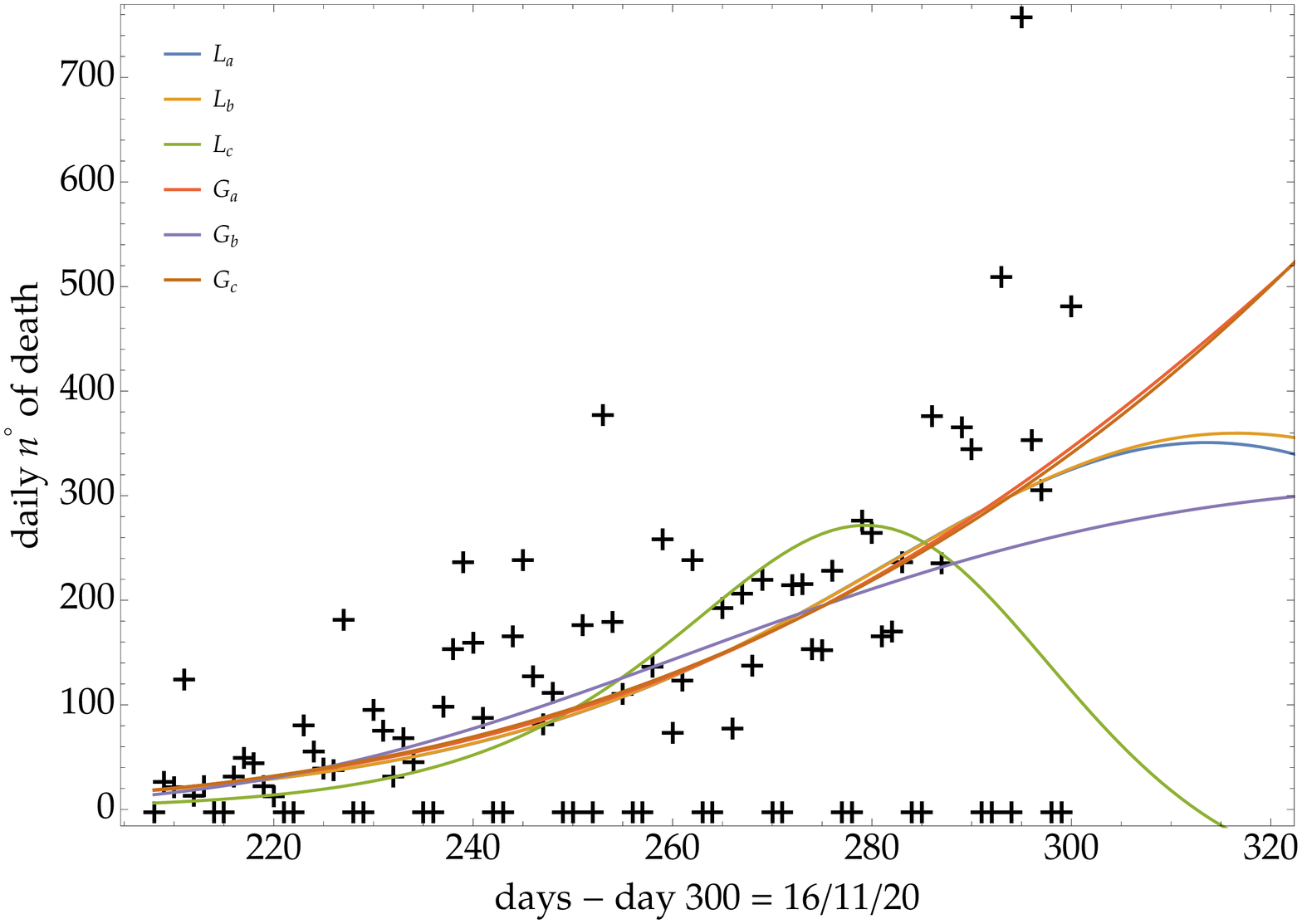}
	\caption{Spain: comparison of the growth laws with the data of the daily number of deaths from the 15th of August to the final day in figure.}
	\label{fig:SpainDday}
\end{figure}

\begin{figure}
	\includegraphics[width=\columnwidth]{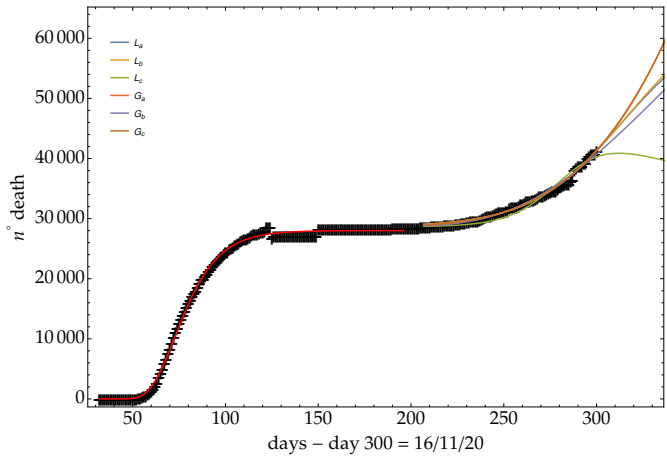}
	\caption{Spain: comparison of the growth laws with the data of the cumulative number of deaths from the initial day 22/Feb to the final day in figure.}
	\label{fig:SpainDtot}
\end{figure}

Figures~\ref{fig:SpainD}, \ref{fig:SpainDday} and~\ref{fig:SpainDtot} report the growth laws results versus  data of the cumulative number of deaths from the 15th of August to the final day, of the daily number of deaths in the same period, and  of the daily number of deaths from the initial day 22/Feb to the final day, respectively.

\section{\label{sec:cc}Comments and Conclusions}

The purpose of this paper is a demonstration-of-concept: one takes simple growth models, considers the available data and shows different scenarios of the future trend of the spreading. 
The method applies a time dependent carrying capacity since the reduction of containment efforts change this crucial parameters of the macroscopic growth laws. 
Different time behaviors of the CC  outline various trends in the new phase.

Therefore a comparison of data, collected in a short time interval, with the plots obtained by the various ansatz for the CC can help to decide if the social isolation conditions have to be strengthened or  weakened. Moreover, a large variation of the CC signals an increase of the pressure on the National Health systems.


\begin{thebibliography}{99}
\bibitem{oms} World Health Organization, Coronavirus disease (COVID-19) outbreak, 
https://www.who.int/emergencies/diseases/novel-coronavirus-2019.

\bibitem{who} World Health Organization, Coronavirus disease (COVID-19) report, https://www.who.int/docs/default-source/coronaviruse/who-china-joint-mission-on-covid-19-final-report.pdf

\bibitem{hopkins}  Novel Coronavirus (COVID-19) Cases, provided by JHU CSSE,  https://github.com/CSSEGISandData/COVID-19.

\bibitem{napoco1} S.~A. Herzog, S.~Blaizot and Niel Hens, Mathematical models used to inform study design or surveillance systems in infectious diseases: a systematic review, BMC Infectious Diseases  \textbf{17} (2017) 775.

\bibitem{napoco2}  N.~C.~Grassly and C.~Fraser, Mathematical models of infectious disease transmission, Nature Reviews Microbiology \textbf{6} (2008) 477.

\bibitem{napoco3}  R. Pastor-Satorras, C. Castellano, P.Van Mieghem and A.Vespignani , Epidemic processes in complex network, Rev. Mod. Phys., VOLUME 87 (2015).

\bibitem{napoco4} P.~Blanchard, G.~F.~Bolz and T.~Kruger, Mathematical modelling on random graphs of sesually trasmitted disease, in Dynamics and Stochastic Process - Theory and Applications, Lecture Notes in Physics, vol. 355, Springer-Verlag, Berlin.

\bibitem{pluc} A.~Pluchino et al.,  
A Novel Methodology for Epidemic Risk Assessment: the case of COVID-19 outbreak in Italy,
	arXiv: 2004.02739

\bibitem{epid1} C. E. Walters, M.I. Mesle', I. M. Hall, Modelling the global spread of diseases: A review of current practice and
capability, Epidemics 25 (2018) 1.

\bibitem{imp} Seth Flaxman, Swapnil Mishra, Axel Gandy et al. Estimating the number of infections and the impact of non- pharmaceutical interventions on COVID- 19 in 11 European countries. Imperial College London (30-03-2020) doi: https://doi.org/10.25561/77731





 N.C. Grassly et al. - COVID- 19 response team - Imperial College 
DOI: https://doi.org/10.25561/78439

\bibitem{istat} L.Fenga, CoViD19: An Automatic,Semiparametric Estimation Method for the Population Infected in Italy, medRxiv
preprint doi: https://doi.org/10.1101/2020.03.14.20036103.

\bibitem{lancet} A.R.Tuite, V. Ng, E.Rees and D.Fisman, Estimation of COVID-19 outbreak size in Italy, Lancet Infect Dis 2020 Published
Online March 19, 2020, doi.org/10.1016/ S1473-3099(20)30227-9

\bibitem{noib}  D.~Lanteri, G.~Carco' and P. Castorina,
How macroscopic laws describe complex dynamics: asymptomatic population and CoviD-19 spreading, in press in International Journal Modern Physics C, arXiv:2003.12457.

\bibitem{gov} Comitato Tecnico Scientifico, Italian Government ( in Italian).

\bibitem{noi1}  P.~Castorina, P.~P.~Delsanto, C.~Guiot, Classification Scheme for Phenomenological Universalities
   in Growth Problems in Physics and Other Sciences, Phys. Rev. Lett. \textbf{96} (2006) 188701.

\bibitem{noi2} P.~Castorina and P.~Blanchard, Unified approach to growth and aging in biological, technical and biotechnical systems, SpringerPlus \textbf{1} (2012) 7.

\bibitem{noia} P.~Castorina, D.~Lanteri and A.~Iorio, Data analysis on Coronavirus spreading by macroscopic growth laws, in press in International Journal Modern Physics C,
arXiv:2003.00507.

\bibitem{gompertz}  B. Gompertz, On the nature of the function expressive of the law of human mortality and a new mode of determining
life contingencies, Phil. Trans. R. Soc. \textbf{115} (1825) 513.

\bibitem{logistic}  P.~F.~Verhulst, Notice sur la loi que la population poursuit dans son accroissement, 
Correspondance Math\'ematique et Physique, \textbf{10} (1838) 113.

\bibitem{review2} P.~S. Meyer, J.~H. Ausubel, Carrying Capacity: A Model with Logistically Varying Limits, Technological Forecasting and Social Change 61 (3): 209-214 1999.

\bibitem{wehldon} T.E.Wheldon, Mathematical Model in Cancer Research, 1988, Adam Hilger ed. 

\bibitem{pop} T. Royama, Analytical Population Dynamics, 1992, Springer ed.

\bibitem{baker} Backer JA, Klinkenberg D, Wallinga J. Incubation period of 2019 novel coronavirus (2019-nCoV) infections among travellers from Wuhan, China, 20-28 January 2020. Eurosurveillance. 2020;25(5). doi:10.2807/1560-7917.ES.2020.25.5.2000062.

\bibitem{guan} Guan W-J et al., Clinical Characteristics of Coronavirus Disease 2019 in China. New Engl J Med. February 2020:NEJMoa2002032. doi:10.1056/NEJMoa2002032.

\bibitem{li} Li Q et al., Early Transmission Dynamics in Wuhan, China, of Novel Coronavirus-Infected Pneumonia. N Engl J Med. 2020;382(13):1199-1207. doi:10.1056/NEJMoa2001316.

\bibitem{lei} Lei S et al. Clinical characteristics and outcomes of patients undergoing surgeries during the incubation period of COVID-19 infection. EClinicalMedicine. April 2020:100331. doi:10.1016/j.eclinm.2020.100331.

\end{thebibliography}
\end{document}